# Direct Determination of Spin-Splitting Energy in Magnetic Graphene by Landau Fan Shifts

Junxiong Hu[1,2,†], Yulei Han[3,4,†], Xiao Chi[1,5,†], Ganesh Ji Omar[1*], Mohammed Mohammed Esmail Alezzi[1], Jian Gou[1], Xiaojiang Yu[5], Rusydi Andrivo[1], Kenji Watanabe[6], Takashi Taniguchi[7], Andrew Thye Shen Wee[1], Zhenhua Qiao[3*], Ariando Ariando[1*]

[1]Department of Physics, National University of Singapore, 117542, Singapore

[2]Centre for Advanced 2D Materials and Graphene Research Centre, National University of Singapore, 117551, Singapore

[3]International Center for Quantum Design of Functional Materials, CAS Key Laboratory of Strongly-Coupled Quantum Matter Physics, and Department of Physics, University of Science and Technology of China, Hefei, Anhui 230026, China

[4]Department of Physics, Fuzhou University, Fuzhou, Fujian 350108, China

[5]Singapore Synchrotron Light Source, National University of Singapore, 5 Research Link, 117603, Singapore

[6]Research Center for Functional Materials, National Institute for Materials Science, Tsukuba, Ibaraki 305-0047, Japan

[7]International Center for Materials Nanoarchitectonics, National Institute for Materials Science, Tsukuba, Ibaraki 305-0044, Japan

†These authors contributed equally to this work.

*Email: ariando@nus.edu.sg; qiao@ustc.edu.cn; ganesh@nus.edu.sg

**Spin-polarized two-dimensional materials with large and tunable spin-splitting energy promise the field of 2D spintronics. While graphene has been a canonical 2D material, its spin properties and tunability are limited. Here, we demonstrate the emergence of robust spin-polarization in graphene with large and tunable spin-splitting energy of up to 132 meV at zero applied magnetic fields. The spin polarization is induced through a magnetic exchange interaction between graphene and the underlying ferrimagnetic oxide insulating layer, $Tm_3Fe_5O_{12}$, as confirmed by its X-ray magnetic circular dichroism. The spin-splitting energies are directly measured and visualized by the shift in their landau fan diagram mapped by analyzing the measured subnikov-de-Haas oscillations as a function of applied electric fields, showing consistent fit with our first-principles and machine learning calculations. Further, the observed spin-splitting energies can be tuned over a broad range between 98 and 166 meV by cooling fields. Our methods and results are applicable to other two-dimensional (magnetic) materials and heterostructures, and offer great potential for developing next-generation spin logic and memory devices.**

# 1. Introduction

Two-dimensional (2D) van der Waals heterostructures comprising various 2D layered materials have emerged as promising building blocks for future ultrafast and low-power electronics. [1,2] Particularly, the integration of 2D non-magnetic and magnetic materials enables efficient spin generation, transport, and modulation by magnetic gates, which is the basis for 2D all-spin logic circuitries. [3-5] To create spin-logic devices, the generation of spin current and long spin transport are the prerequisites. Graphene is an ideal spin channel due to its long spin diffusion length and long spin lifetime even at room temperature. [6,7] Although pristine graphene is not intrinsically spin-polarized, magnetic proximity effect (MPE) can be used to induce spin splitting in graphene by coupling with magnetic substrates, leading to spin-dependent transport in magnetic graphene. [8-10] The proximity-induced spin splitting energy, originating from the electron wavefunction overlap between two adjacent materials, is a result of interface exchange interaction[11-13] and has been observed through Zeeman spin Hall effect, [8,10] anomalous Hall effect, [14,15] and Hanle precession measurements.[4,9,16] However, none of these previous experiments could directly quantify spin-splitting energy, which determines the spin-dependent density of charge carriers. [4] Moreover, the large discrepancy between theoretical predictions and experimental determination of spin-splitting energy still needs to be resolved (**See summary in Table S1**).

The goal of MPE is to achieve a considerable spin-splitting energy while maintaining the exceptional capabilities of spin transport in magnetic graphene. [1] Theoretically, spin splitting can be enhanced by reducing the separation between the graphene sheet and magnetic substrate. [11,12] However, experimentally it is highly challenging to achieve this, for example, by applying external pressure. Previously, a field cooling technique has been used to enhance the spin-splitting energy in graphene/antiferromagnet CrSe heterostructure.[17] The magnetic order at the interface can be modulated by the cooling fields, enabling the enhancement of the spin splitting energy. However, the CrSe magnetic substrate is conducting and inevitably short-circuits the graphene layer, making it difficult to disentangle the true origin of the effects from the substrate. Therefore, spin-splitting enhanced by a cooling field begs for a further direct experimental demonstration that can be achieved by using a magnetic insulator,[18] as demonstrated in our work.

In this article, we directly quantify spin-splitting energy in magnetic graphene using Landau fan shifts attained from the electric-field tuning of the Subnikov-de-Haas oscillation frequency. By coupling with a ferrimagnetic insulator, $Tm_3Fe_5O_{12}$ (TmIG), we observe a strong spin splitting energy of 132 meV at 2 K. First-principles calculations reveal that TmIG induces an electron doping to graphene, and the proximity-induced spin splitting energy can be up to 133 meV, consistent with that obtained from Landau fan shifts. Moreover, the spin splitting is visualized by the color map of the Landau fan diagram. Furthermore, we demonstrate that the cooling field technique can be applied to our graphene/TmIG heterostructure, leading to the tuning of the spin splitting energy between 98 and 166 meV. This result is supported by our machine learning fitting based on a phenomenological model. Finally, the spin polarization of graphene π orbitals is probed directly using X-ray magnetic circular dichroism and is confirmed to originate from the carbon edges.

## 2. Results and discussion

### 2.1 Concept of Landau fan shift

Graphene exhibits a Dirac-cone low-energy spectrum with linear energy dispersion, protected by both inversion and time-reversal symmetries, and thus the low-energy quasiparticles in graphene behave as massless Dirac fermions.[19] Under a perpendicular magnetic field $B$, the resistivity of the Dirac fermion can show quantum oscillations associated with its discrete dispersionless Landau-levels (LL) that can be described by a sequence function, $E_N$, of the square-root of field and LL-index $N$ as $E_N = \text{sgn}(N)\sqrt{\alpha|N|B}$, where $\alpha = 2e\hbar v_F^2$ is a constant with $e$ the electronic charge, $\hbar$ Planck constant, and $v_F$ Fermi velocity (**Supplementary Note 1**).[20,21] For pristine graphene, each LL has a four-fold of double-spin and double-valley degeneracy.[22,23] At each fixed Fermi energy $E_F$, the discrete set of $N$ aligns linearly on a fan of lines $N = \pm\frac{E_F^2}{\alpha B} \pm \frac{1}{2}$, where $\pm\frac{1}{2}$ comes from the zeroth LL, forming the so-called Landau fan diagram. In pristine graphene, the slope $\beta$ of this Landau fan diagram is only determined by Fermi energy following (**Supplementary Note 2**):

$$\beta = \pm \frac{E_F^2}{\alpha} \tag{1}$$

The oscillation frequency $B_F$, derived from the magnitude of $\beta$, is thus also a function of Fermi energy. For pristine graphene, the linear extrapolation of carrier-density dependence of $B_F$ for electrons and holes will cross $B_F$ at zero at Dirac point (**Fig. 1a**).

In proximity to a magnetic substrate, graphene's time-reversal symmetry can be broken, which can lift its spin-degeneracy (*i.e.*, graphene is spin-polarized).[11-13] As a consequence, LLs of spin-polarized graphene will be modified by a spin splitting term: $E_N^{\uparrow,\downarrow} = \text{sgn}(N^{\uparrow,\downarrow})\sqrt{\alpha B} \times \sqrt{|N^{\uparrow,\downarrow}|} \pm \frac{\Delta}{2}$, where $N^{\uparrow,\downarrow}$ is the index for spin-up and spin-down LLs and $\Delta$ is the spin splitting energy (**Supplementary Note 1**). For spin-polarized graphene, the energy spectrum will shift up and down by $\Delta/2$ for spin-up and spin-down LLs, respectively. Meanwhile, the slope $\beta$ of the Landau fan diagram will also be modified as (**Supplementary Note 2**):

$$\beta = \pm \left(\frac{2E_F^2}{\alpha} + \frac{\Delta^2}{2\alpha}\right) \tag{2}$$

Similar to the case of pristine graphene, the $B_F$ of spin-polarized graphene is also a function of Fermi energy but with three distinct features: First, it depends on $E_F$ and $\Delta$ that can be tuned independently. Second, it increases twice as fast ($\sim 2E_F^2/\alpha$), compared to that of pristine graphene ($\sim E_F^2/\alpha$). Third, the linear extrapolation of its carrier-density dependence for electrons and holes crosses $B_F$ at a non-zero value of $\Delta^2/2\alpha$ at Dirac point (**Fig. 1b**), in stark contrast to that of pristine graphene, which crosses at zero (**Fig. 1a**). Therefore, we can use this non-zero intercept to evaluate the spin-splitting energy $\Delta$ in magnetic graphene.

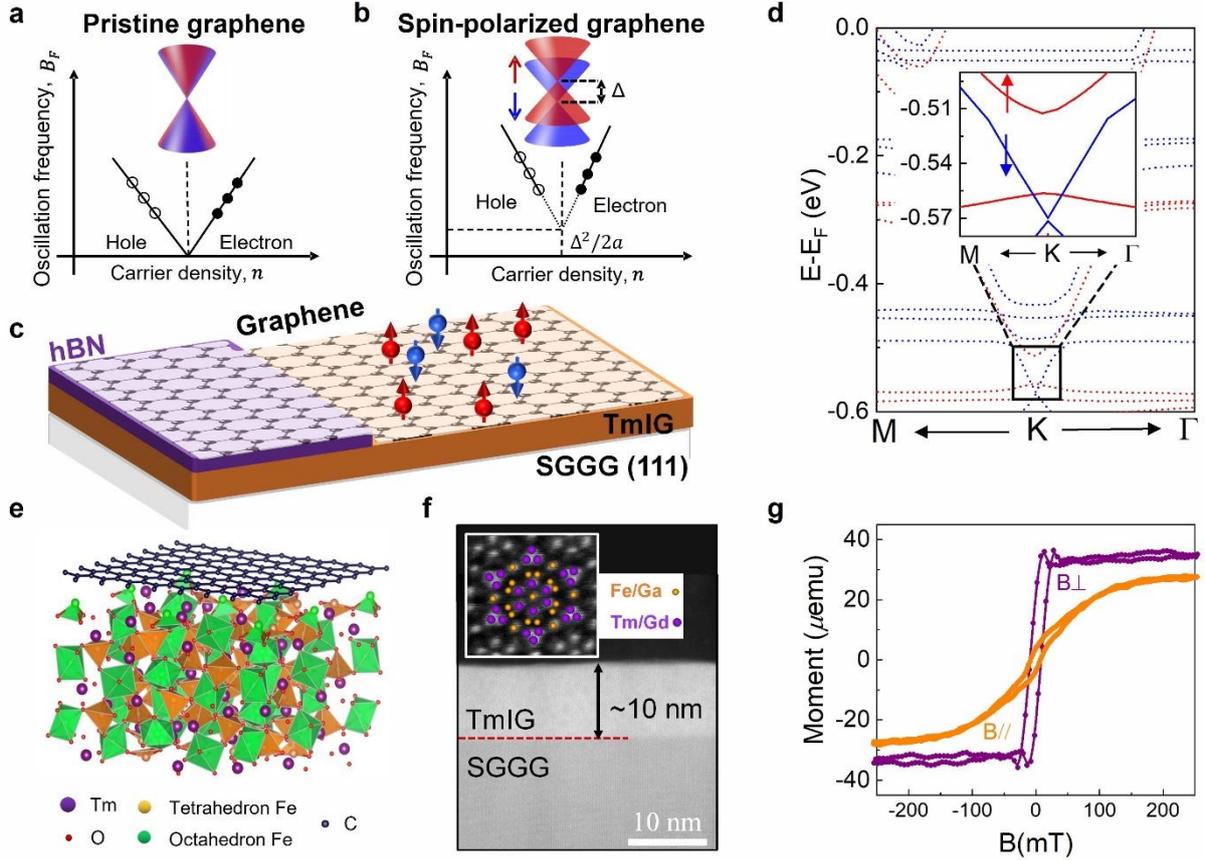

**Figure 1.** Characterizations and spin splitting in graphene (G) on $Tm_3Fe_5O_{12}$ (TmIG). Schematic of oscillation frequency $B_F$ as a function of carrier density $n$ a) for pristine graphene (The linear fits of electrons and holes extrapolate to zero-intercept at charge neutral point), and b) for spin-polarized graphene (In the region of $|E_F| > \frac{\Delta}{2}$, the linear fits of electrons and holes extrapolate to a non-zero intercept at charge neutral point). c) Schematic of G/TmIG heterostructure. The hBN is used to separate the graphene and TmIG as a contrast study. d) Calculated spin-resolved band structure of G/TmIG heterostructure. Bands in red (blue) correspond to spin up (down). The inset shows the enlargement of the calculated low energy bands at K point. e) Supercell for graphene on a $Tm_3Fe_5O_{12}(111)$ surface. Blue (the top layer), purple (large), yellow and green (medium), and red (small) colors represent C, Tm, Fe, and O atoms, respectively. f) Cross-sectional STEM image for a TmIG/sGGG(111) sample viewed along the [1$\bar{1}$0] direction. The red line marks the boundary between TmIG and sGGG. The inset shows the high-magnification plan-view STEM image with an overlay of the [111]-projection of TmIG lattice. g) Magnetic hysteresis loops in perpendicular and in-plane magnetic fields of TmIG, showing the out-of-plane magnetic anisotropy.

## 2.2 Device design and characterization

As a proof of concept, we designed graphene/$Tm_3Fe_5O_{12}$ (G/TmIG) heterostructures, as shown in **Fig. 1c**. TmIG was chosen to be the magnetic substrate for the following reasons: First, it is insulating so that graphene provides the only transport channel, making the tuning of Fermi energy possible;[24] Second, it has a perpendicular magnetic anisotropy, and its orbital hybridizes with graphene sufficiently to induce a strong exchange coupling;[25,26] and third, it has a low Gilbert damping and a high Curie temperature of 560 K, promising for a high-temperature magnetic exchange interaction.[27] As a result, the unbalanced spin-up and spin-down electrons due to the lifting of the spin degeneracy will lead to spin-dependent quantum transport. **Figure 1.1e** shows a supercell used in the atomic and electronic structure calculation of graphene on a (111) surface of TmIG. There are two sites of Fe atoms in the TmIG unit cell: $Fe^{3+}$ octahedral ($Fe^{3+}$ Oh), two per formula unit and $Fe^{3+}$ tetrahedral ($Fe^{3+}$ Td), three per formula unit. The $Fe^{3+}$ Td atoms are coupled antiferromagnetically to those $Fe^{3+}$ Oh, but have a remanent magnetic moment.[28] **Figure 1d** shows the optimized spin-polarized band structure of G/TmIG heterostructure. The Dirac cones are below the Fermi level and located inside the bandgap of TmIG. The inset zooms in the spin-polarized π bands near the Dirac point at K and shows that the induced spin splitting energy is ~60 meV at an interlayer distance of 2.5 Å. Since the spin splitting is induced by the interaction between graphene and TmIG, the strength can be further enhanced by decreasing the separation between graphene and TmIG.[14] For example, the spin splitting can be up to ~133 meV at $d = 2.0$ Å under certain pressure, indicating a strong exchange interaction (**Fig. S1**).

High-quality TmIG thin films were epitaxially grown on (111)-oriented substituted gadolinium gallium garnet (SGGG) substrates by pulsed laser deposition (PLD) (**See Methods**). **Figure 1f** shows a cross-sectional scanning tunneling electron microscopy (STEM) image of 10 nm TmIG coherently grown on SGGG without obvious dislocations with the red dashed line indicating TmIG/SGGG interface. The atomic arrangements of concentric hexagons are displayed in the high-magnification image, exhibiting a remarkable snowflake pattern (Inset of **Fig. 1f**) (**Fig. S2**). Because of the negative magnetostriction constant of TmIG, the interfacial tensile strain exerted by SGGG produces a perpendicular magnetic anisotropy in TmIG.[25,26] The magnetic hysteresis loops shown in **Fig. 1g** confirm the out-of-plane magnetic anisotropy with an out-of-plane coercive field of a few oersteds and a saturation field of <20 Oe, while the in-plane loop indicates a typical hard-axis behavior with a saturation field of ∼150 Oe. Our device structure is illustrated in **Fig. 2b**. We place graphene on TmIG (also labelled as B-TmIG for clarity) using a PMMA/PMGI dry transfer technique (**See Methods**).[29] A top hBN (T-hBN) layer was then transferred to cover the entire graphene region, which simultaneously protects graphene from chemical contaminations and serves as a top-gate dielectric layer.[30] At a section of the same sample, a 20-nm hBN buffer layer was inserted to screen the exchange interaction between graphene and TmIG as a contrast study (**Fig. S3**). Finally, a standard electron lithographic technique was used to create Hall bar geometry for electronic transport measurements. We have studied several devices and focused on two of them (device D1 and D2) showing representative transport behavior.

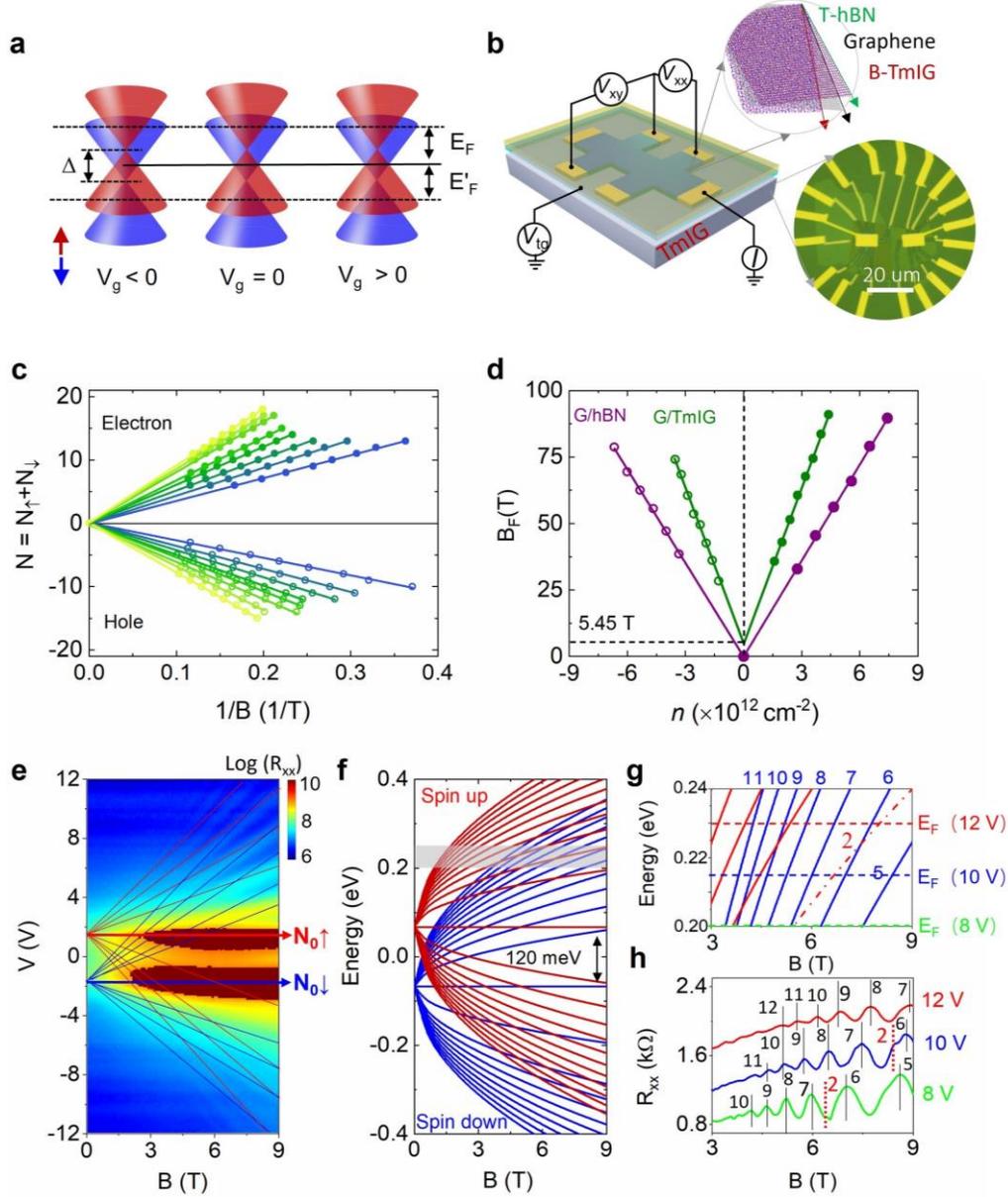

**Figure 2.** Quantum oscillations tuned by gate voltages in graphene on TmIG. a) Schematic of the gate-tunable Fermi energy in spin-polarized graphene. The red and blue cones are the linear dispersion near the Dirac points corresponding to spin-up and spin-down electrons. b) schematic of top-gate device and the art view of T-hBN/graphene/B-TmIG structure. The bottom right is the optical image of device. c) Landau fan diagram of the LL index $N$ versus $1/B$ for gate voltages from ±8 to ±22 V (from bule to yellow). Each data point represents the value of $1/B$, where the $N$th minimum of $R_{xx}$ occurs. The lines correspond to a linear fit with the slope indicating the $B_F$. d) The dependence of $B_F$ on the carrier density $n$ for G/TmIG and G/hBN. The positive $n$ corresponds to electrons and negative to holes. e) A spin splitting Landau fan diagram for G/TmIG. Longitudinal resistance $R_{xx}$ versus magnetic field and gate voltage. f) Theoretical calculation of Landau fan spectrum with a given spin splitting energy. The split LL spectrum comprises up (red) and down (blue) spins. The position of $E_F$ is denoted by the dark shadow area. g) Zooming-in LL states between 3 - 9 T and $E_F$ = 0.20-0.24 eV from f. The red dashed line highlights the spin-up LL $N\uparrow$ = 2. h, The measured SdH oscillations at gate voltages of 8, 10 and 12 V. Data in panels c-d are from device D1; Panels e-h from device D2.

**2.3 Determination of spin-splitting energy**

First, we study gate-voltage tuning of Shubnikov-de Haas (SdH) oscillations at a constant spin splitting energy of G/TmIG heterostructure (**Fig. 2a**). Similar to graphene on silicon dioxide,[31] the electron and hole carriers in G/TmIG can also be tuned by gate voltages. The gate voltages tune the Fermi energy, leading to the modulation of oscillation frequency,[22,23] as shown in **Fig. S4** for gate voltages from $\pm 8$ to $\pm 22$ V. In order to obtain $B_F$, we extract LL indices by assigning consecutive integers to each SdH dip. Because $N \to 0$ has to be satisfied for $1/B \to 0$, there is only one way to assign the integer values of $N$ to the SdH dips (**Fig. S5**).[17] Based on this, the sequence of values of $1/B_N$ is then plotted against $N$ indices, showing a linear fan diagram (**Fig. 2c**). The $B_F$ can then be derived from the integer change in the slopes of this fan diagram, and the dependence of $B_F$ on the carrier density $n$ is shown in **Fig. 2d**. For G/TmIG heterostructure, the electron and hole carriers both exhibit a linear dependence $B_F = c_1 n + b$, with $c_1 \approx 1.98 \times 10^{-15}$ Tm$^2$ (± 2%) and $b$ = 5.45 T (± 6%), where ± refers to the standard error for linear fitting. As can be seen in **Fig. 2d**, a distinct feature is that while the linear extrapolation of $B_F$ has a zero intercept at Dirac point for G/hBN heterostructure (**Fig. S6**), it is extrapolated toward a non-zero intercept ($B_F = 5.45$ T) for the G/TmIG heterostructure, consistent with the model discussed in **Figs. 1a** and **1b**. Using the $B_F$ intercept of 5.45 T, we can quantify Δ in device D1 to be 132 meV at 2 K, in good agreement with the theoretical calculation based on the split Landau level spectra (**Fig. S7**). Moreover, the Hall plateau value also provides further verification of the predicted LL structure with a given spin-splitting, calculated by appropriately summing the Chern numbers of the LLs from the spin-up and down subbands (**Fig. S8**).

Since each peak in the SdH oscillation corresponds to a crossing between one Landau level and Fermi level, we can examine the signature of spin splitting by comparing the measured color mapping of Landau fan diagram to the splitting in the Landau level spectrum (**Figs. 2e-h**). Another distinct feature can be seen in the color maps of the Landau fan diagram. For G/hBN, there is a single conventional Landau fan diagram, where the trajectory of each quantum oscillation evolves in a straight line away from the CNP at zero fields (**Fig. S9**). In stark contrast with G/hBN, G/TmIG shows a splitting Landau fan diagram, where the entire sequence of spin-up and spin-down LL crossings can be seen in the map (**Fig. 2e).** The Landau fan diagram splits into two sets, sitting at the two sides of the zero-gate voltage. The split trajectory of each quantum oscillation evolves in a straight line away from the two sides. This feature is absent in the Landau fan diagram of G/hBN since no spin splitting is involved. For the split Landau diagram, the LLs of graphene are modified by a spin-splitting term: $E_N^{\uparrow,\downarrow} = \text{sgn}(N^{\uparrow,\downarrow})\sqrt{\alpha B} \times \sqrt{|N^{\uparrow,\downarrow}|} \pm \frac{\Delta}{2}$, where $N^{\uparrow,\downarrow}$ is the index for spin-up and spin-down LLs and $\Delta$ is the spin-splitting energy, as drawn in **Fig. 2f**. Compared with this theoretical Landau fan spectrum, the two sets of Landau energy levels are spin-up and spin-down diagram. At positive $V_{TOP}$ = +1.5 V, well-defined LLs are seen as part of a Landau fan from LLs with N = 0, 1, 2…. At negative $V_{TOP}$ = -1.5 V, LLs emerge from a second Landau fan. The split two sets of the Landau fan diagram sit at the two sides of the zero-gate voltage associated with the spin-up and spin-down spectrum. Therefore, the full-color map of the Landau fan diagram can visualize the spin splitting in G/TmIG. Second, by zooming-in the Landau-level crossing between 3 - 9 T (**Fig. 2g**), there is a spin-up LL N↑= 2 (red dashed line) located between spin-down LL N↓= 6 and N↓= 7 at Fermi energy of 0.20 eV. With the increase of Fermi level, the spin-up LL N↑= 2 will shift right and finally cross N↓= 6 towards N↓= 5. This physical picture is consistent with the measured SdH oscillations (**Fig. 2h**). At a gate voltage of 8 V ($E_F$ = 0.20 eV), there is a shoulder peak belonging to N↑=2 located between N↓= 6 and N↓= 7. With the gate voltage increases to 10 V ($E_F$ = 0.215 eV), the shoulder peak shifts right and finally disappear at a gate voltage of 12 V ($E_F$ = 0.23 eV), consistent with the physical picture in **Fig. 2g**. While for N↑ = 3, the Landau levels are dense and therefore cannot be distinguished by $R_{xx}$. Therefore, by using the combined split LL spectra and measured color map of the Landau fan diagram, we can conclude that the spin splitting is induced in G/TmIG heterostructure that can be tuned by a gate voltage.

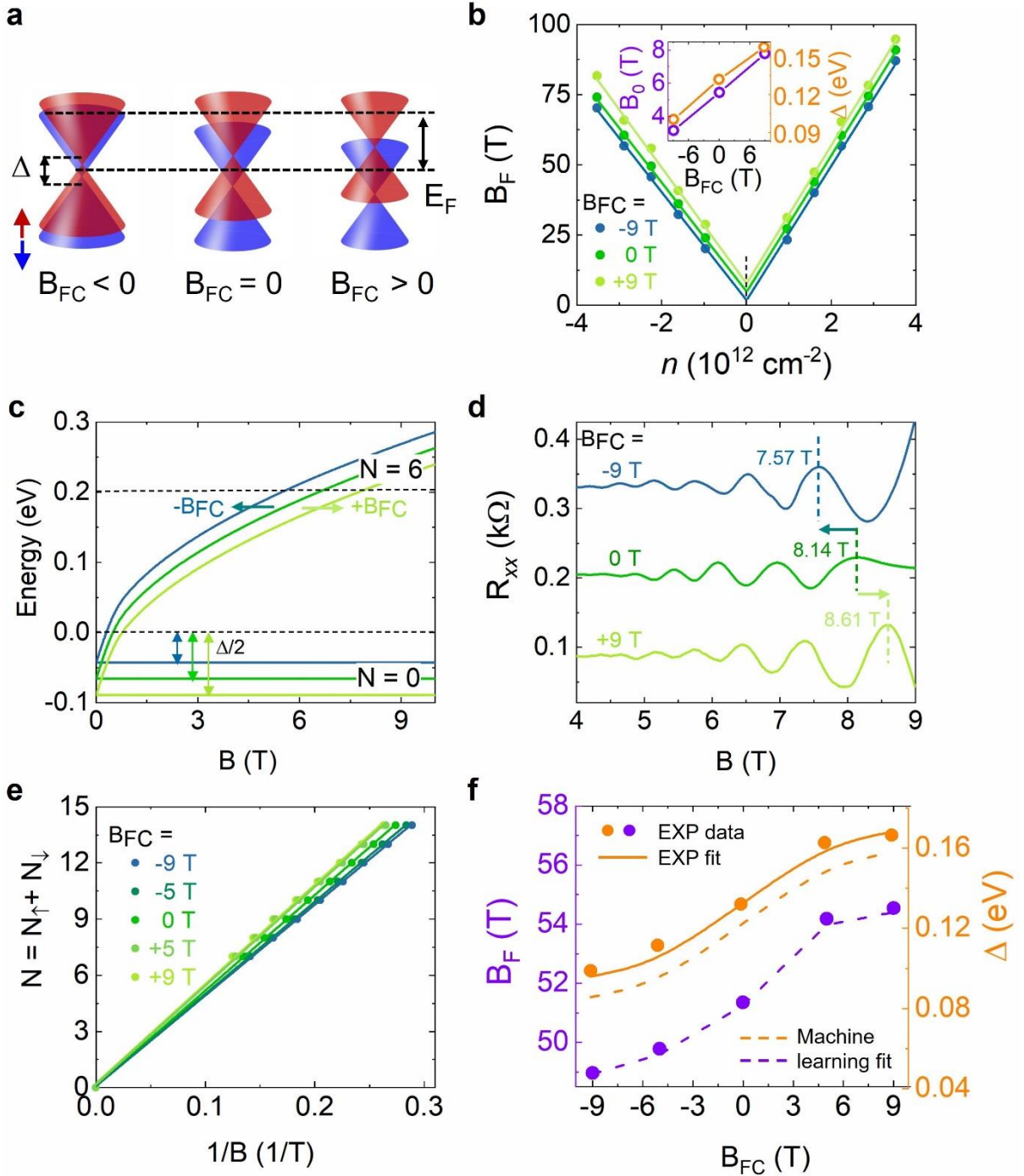

**Figure 3.** Quantum oscillations tuned by cooling fields in graphene on TmIG. a) Schematic for the spin splitting energy tuned by cooling fields in spin-polarized graphene. b) The dependence of $B_F$ on the carrier density $n$ at cooling fields of 0 and ±9 T. The inset shows the dependence of intercept $B_0$ and the corresponding spin-splitting energy on the cooling fields. c) Illustration of spin-down LL N↓ = 6 shifted by cooling fields. d) The shift of quantum oscillations by cooling fields at the gate voltage of 12 V. The peaks of N↓ = 6 are marked by the dashed lines. e) LLs indices N are extracted in the Landau fan diagram for different cooling fields, showing a linear relationship. f) The oscillation frequency and the spin splitting energy tuned by cooling fields. The solid dots and lines are experimental data, and the dashed lines are model fitting by a machine learning method. Data in panels b-f are from device D1.

## 2.4 Cooling-field tuning of spin-splitting energy

Next, we study the cooling-field tuning of the SdH oscillations at a constant Fermi energy in G/TmIG heterostructure (**Fig. 3a**). By applying a magnetic field perpendicular to the sample plane during cooling (namely cooling field, $B_{FC}$), a remnant out-of-plane magnetic moment can be produced in the heterostructures. This field-cooling technique provides another knob that can be used to tune the splitting energy of spin-polarized G/TmIG heterostructures by flipping the applied magnetic field direction, where a positive cooling field results in an increase while a negative cooling field results in a decrease of the spin splitting energy,[17] as illustrated in **Fig. 3a**. Since the intercept of $B_F$ is also directly related to the spin-splitting energy (**Fig. 1b**), the modulation of this spin-splitting energy by cooling fields can then be revealed from the intercept of $B_F$ at Dirac point, $B_0$ (**Fig. 3b**). Without a cooling field, the intercept $B_0$ is 5.45 T, corresponding to a spin splitting energy of 132 meV. By applying a negative cooling field of -9 T, the intercept $B_0$ decreases from 5.45 to 3.06 T, indicating that the spin splitting energy decreases from 132 to 98 meV. While using a positive cooling field of +9 T, the intercept $B_0$ increases to 8.65 T, indicating that the spin splitting energy increases to 166 meV (**Fig. S10**). Therefore, a cooling field of up to 9 T can then be used to tune the spin splitting energy between 98 and 166 meV in G/TmIG (**Inset of Fig. 3b**).

**Figure 3c** illustrates the energy shift of LLs by cooling fields. By applying a positive cooling field, the increase of the spin splitting energy right-shifts the LL $N\downarrow = 6$, while for a negative cooling field, the decrease of the spin splitting energy shifts the LL $N\downarrow = 6$ to the opposite direction. In order to study the tunability of spin-splitting energy by cooling fields, we set the Fermi energy at a constant of 0.19 eV by applying a top gate voltage $V_{Top} = 12$ V. The shift of the energy spectrum can then be observed using SdH oscillations. As can be seen from **Fig. 3d**, at 12 V, the peak position of $N\downarrow = 6$ is ~8.14 T at zero cooling field. Applying a positive cooling field of +9 T right-shifts this peak to 8.61 T, while applying a negative cooling field of -9 T left-shifts the peak to 7.57 T, consistent with the physical picture in **Fig. 3c**. On the other hand, no obvious shift was observed for G/hBN heterostructure, indicating that the exchange interaction between graphene and TmIG has been screened by hBN (**Fig. S11).** Moreover, the shifts in the Hall signal of G/TmIG are also consistent with the shifts in the SdH oscillations, *i.e.*, the Hall plateau right-shifts with the positive cooling field and left-shifts with the negative cooling field (**Fig. S12**). These transport studies suggest that the band structure of graphene shifts with $E_F$ because of the spin splitting, which can be tuned by the cooling fields.

To further study the field-cooling dependence of $B_F$, we also plot $R_{xx}$ as a function of 1/B for different cooling fields, from which discrete sets of LL-indices $N$ linearly proportional to 1/B were obtained with slopes of $\beta$ for different cooling fields (**Fig. 3e**). According to the slopes of the Landau fan diagram, the $B_F$ tuned by cooling fields has a range of 48.6 - 54.3 T. Since the Fermi energy is 0.19 eV at 12 V, it can be deduced that the spin splitting energy under the field cooling of ±9 T can be tuned between 98 and 166 meV (**Fig. 3f**). To examine the induced spin-splitting energy in G/TmIG, we then used a machine learning method to extract $E_F$ and $\Delta$ by fitting the experimental data of $B_F$, which automatically generate the values of Fermi energy and spin splitting energy by minimizing the loss function (**See Methods**). By solving a standard machine learning problem, these four parameters can be obtained: $E_F$ = 0.193 eV, $\Delta_0$ = 0.122 eV, $\Delta_1$ = 0.038 eV and $\lambda$ = 0.21 T$^{-1}$ (**Fig. S13**). The machine learning fitting shows that in G/TmIG heterostructure, the Fermi energy is 0.193 eV at 12 V, which is quite close to our experimental data of 0.19 eV (0.2% error). Moreover, the spin splitting energy is 122±36 meV (dashed lines in **Fig. 3f**), where 122 meV is the spin splitting energy without cooling field, close to our experimental data of 132 meV (7.6% error). These results emphasize that spin splitting is indeed induced in graphene on TmIG, and the cooling fields can be used to tune the spin splitting energy.

## 2.5 XMCD measurement of the spin polarization

Finally, in order to directly identify the origin of the proximity-induced magnetism in graphene by TmIG, we applied elemental-specific techniques, synchrotron radiation-based X-ray absorption spectroscopy (XAS) and X-ray magnetic circular dichroism (XMCD).[32] These measurements uniquely provide the origin of the magnetic momentum and coupling at the interface. The detailed sample structure and experimental configuration are shown in **Fig. S14**. **Figure 4a** shows two pronounced regions 283–289 eV and 289–315 eV, which are ascribed to the transitions of core electrons into $\pi^*$ (C 1s→ $\pi^*$) and $\sigma^*$ orbitals (C 1s→ $\sigma^*$), respectively. Compared with pure graphite in the transitions of C 1s→ $\pi^*$ and C 1s→ $\sigma^*$, the spectra collected for G/TmIG become broadened and shifted to higher photon energy (~0.1 eV), which can be attributed to the orbital hybridization at the interface between graphene and TmIG.[33] **Figure 4b** further shows the angle-dependent XAS spectra at C K edge. The spectral feature around 285 eV, which is assigned to the transitions of C 1s→ $\pi^*$, is significantly enhanced when X-ray is at grazing incidence (30 degrees) and vanishes for increasing incident angle, indicating that the $\pi^*$ unoccupied states of C $2P_Z$ orbitals are aligned out-of-plane in the sp$^2$ coordination.

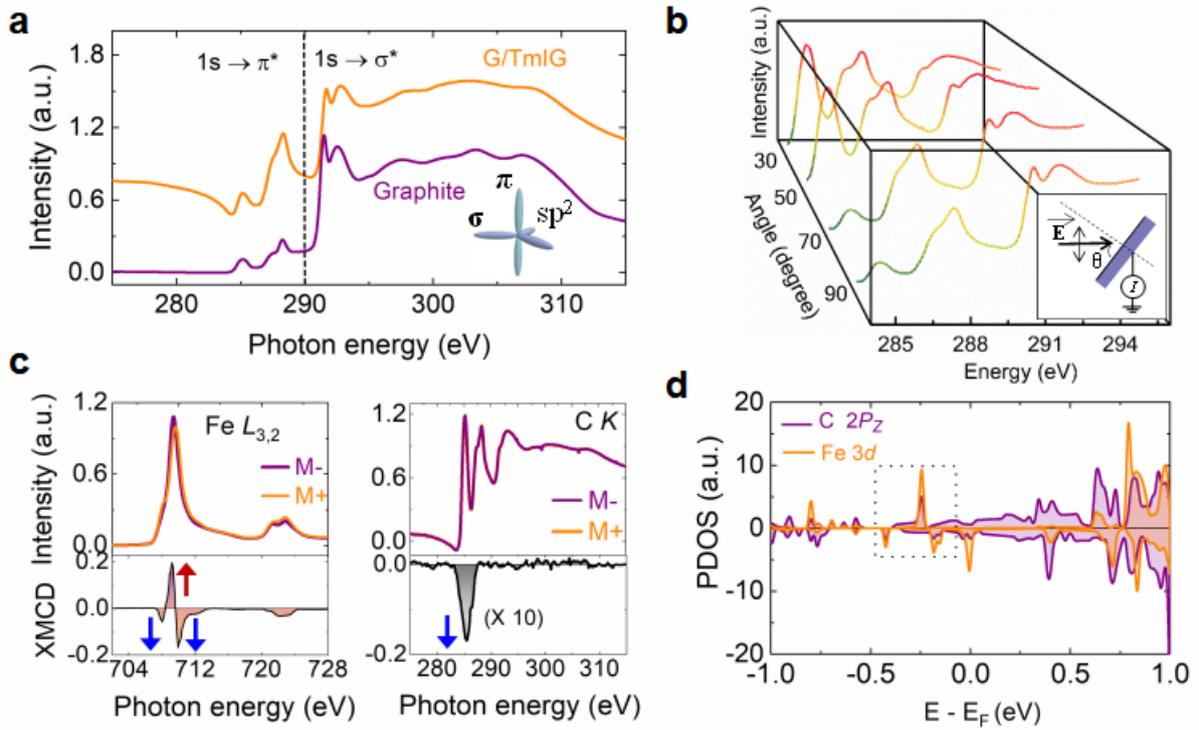

**Figure 4.** XAS and XMCD of C *K*-edge and Fe *L*-edge. a) XAS acquired at the carbon *K*-edge of the G/TmIG. The inset shows the *sp²* hybrid orbitals: One *s* orbital and two *p* orbitals form a σ bond with trigonal planar structure. The half-filled $P_Z$ orbitals, perpendicular to the planar structure, form the π bonding and π* antibonding band touching at the corner of the hexagonal Brillouin zone. b) Angle-dependent C *K*- edge spectra of the G/TmIG. The incident angle is the angle between the electrical vector of the light and the normal of the surface. c) XMCD spectra of the G/TmIG system: Absorption spectra of the Fe $L_{2,3}$- and C *K*-edges are measured at ±2T magnetic field. The corresponding XMCD signals are shown in the lower panels of the respective figures. d) The partial DOSs of Fe 3d orbitals and C $2P_Z$ orbitals. The dashed square box highlights the overlaps of orbitals.

**Figure 4c** shows XMCD spectra of the G/TmIG system. The Fe $L_{3,2}$ XAS and XMCD spectrum, which shows positive and negative peaks corresponding to tetrahedral (Tet $Fe^{3+}$) and octahedral (Oct $Fe^{3+}$) sites, respectively, which have antiparallel coupled spins.[34] The most important finding is the observation of the obvious XMCD contrast at the C K absorption edge. The C K spectrum has the maximum asymmetry at 285 eV, indicating an induced magnetism in graphene, and also that the magnetism mainly comes from the spin-polarized C $2P_Z$ orbitals. [32,33] Moreover, from the negative sign of the XMCD signal, we can conclude that the orbital moment of graphene is aligned parallel to the moment of the Oct $Fe^{3+}$, which is consistent with our first-principles calculations. As shown in **Fig. 4d**, the partial density of state (PDOS) shows an overlap of Fe 3d and C $2P_Z$ orbitals, highlighted by the dashed square box, suggesting that the C $2P_Z$ orbitals of graphene are spin polarized due to the hybridization of the Fe 3d and the C-pz orbitals. Therefore, by combining XMCD and first-principles calculations, we can conclude that the hybridization between graphene π band and Fe 3d band leads to partial charge transfer from TmIG substrate to the spin-polarized electronic states of graphene, leading to the induced magnetism in graphene.

## 3. Conclusion and outlook

We have demonstrated that the insulating TmIG with perpendicular magnetic anisotropy can hybridize with graphene sufficiently to induce a strong spin-splitting energy. The present measurement concept based on Landau fan shifts allows us to directly quantify the spin splitting energy, which is supported by our first-principles calculations and machine learning fitting. The combination of direct determination of spin-splitting energy and the theoretical calculations may also help reduce the discrepancies between theoretical predictions and experimental findings in other magnetic graphene heterostructures, for example, graphene/ $Y_3Fe_5O_{12}$ heterostructure.[14,16,18] However, we note that spin-splitting energy in magnetic graphene might not only be affected by the proximity strength between the two adjacent materials but also by interfacial quality such as roughness, defects, dangling bonds, and chemical contaminants. This interface quality could also lead to the observed discrepancy between theoretical prediction and experimental findings. Therefore, the development of dry transfer in our work or direct large-scale growth of graphene on a magnetic substrate can ensure high-quality interface,[8,15] maximizing the spin splitting energy in magnetic graphene, and finally leading to substantial advances in 2D spintronics with a long-distance transfer of spin information and ultrafast operation.

**Acknowledgements**

This work is supported by the Ministry of Education (MOE) Singapore under the Academic Research Fund Tier 2 (Grant No. MOE-T2EP50120-0015), by the Agency for Science, Technology and Research (A*STAR) under its Advanced Manufacturing and Engineering (AME) Individual Research Grant (IRG) (Grant No. A2083c0054), and by the National Research Foundation (NRF) of Singapore under its NRF-ISF joint program (Grant No. NRF2020-NRF-ISF004-3518). The work at USTC is financially supported by the National Natural Science Foundation of China (Grants No. 11974327 and No. 12004369), Fundamental Research Funds for the Central Universities (Grants No. WK3510000010 and No. WK2030020032), Anhui Initiative in Quantum Information Technologies. We also thank the supercomputing service of AM-HPC and the Supercomputing Center of the USTC for providing the high-performance computing resources. K.W. and T.T. acknowledge support from the Elemental Strategy Initiative conducted by the MEXT, Japan (Grant Number

JPMXP0112101001) and JSPS KAKENHI (Grant Numbers 19H05790, 20H00354 and 21H05233).

**Author contributions**

A. A. conceived and supervised the project. J. X. H. designed and performed the experiments. Y. L. H. carried out first-principles calculations and machine learning fitting under the supervision of Z. H. Q. X. C. and X. J. Y. helped with the XMCD measurements. G. J. O. prepared the TmIG films. M. M. A. E. helped with the theory of splitting Landau fan diagram. T. T. and K. W. provided the bulk hBN crystals. J. G. and A. T. S. W. helped in data analysis and interpretation. J. X. H., G. J. O. and A. A. analyzed the experimental data and wrote the manuscript. All authors discussed the results and commented on the manuscript.

**Competing interests**

The authors declare no competing interests.

**Additional information**

Supplementary information is available for this paper at.

Reprints and permissions information is available at.

Correspondence should be addressed to A. Ariando.

**Data availability.** The data that support the plots within this paper and other findings of this study are available from the corresponding authors upon reasonable request.